# STUDY OF A HYBRID - ANALOG TV AND ETHERNET- HOME DATA LINK USING A COAXIAL CABLE

Radu ARSINTE (*)
*(*) Technical University Cluj-Napoca , Tel: +40-264-595699, Str. Baritiu 26-28, Radu.Arsinte@com.utcluj.ro*

**Abstract:** The paper presents an implementation and compatibility tests of a simple home network implemented in a non-conventional manner using a CATV coaxial cable. Reusing the cable, normally designated to supply RF modulated TV signals from cable TV networks, makes possible to add data services as well. A short presentation of the technology is given with an investigation of the main performances obtained using this technique. The measurements revealed that this simple solution makes possible to have both TV and data services with performances close to traditional home data services: cable modems or ADSL, with minimal investments. This technology keeps also open the possibility for future improvements of the network: DVB-C or Data via Cable Modems.

## I. IN-HOME TV AND DATA DISTRIBUTION

Cable-based services are defined as application services that are delivered via a hybrid fiber/coax (HFC)/cable infrastructure. Cable operators currently offer a wide variety of cable-based services; additional service opportunities are enabled by the advent of home networks. Examples of such services include high-speed data, streaming audio and video, packetized telephony, network management, home security, environmental monitoring, medical monitoring, gaming, interactive television, and video conferencing.

Internet has become a key player for the last years in the multimedia market, and will continue to enhance its importance. With producers (companies) and consumers (buyer) going online, the whole market is going through a paradigm shift. Digital television will bring a new era to the market, but there is also no doubt that Internet is there to continue the trend. With digital television being network enabled, the next shift is whether the PC (personal computer) or the digital television will take the lead, or will they converge to form a new generation product. Adaptation of the new home to the future scenario is necessary; choosing the right architecture for TV and data distribution plays a major role.

Most of the Cable TV distribution is realized using broadband coaxial cable. In modern systems only the last mile is implemented using copper cable, the rest of network is based on optical fiber.

Using optical fiber as a distribution media inside the home is related with major difficulties, optical connections and the related equipment being too expensive for domestic purposes.

This domestic network is a part o a larger architecture. As an example, in figure 1 a cable network is presented.

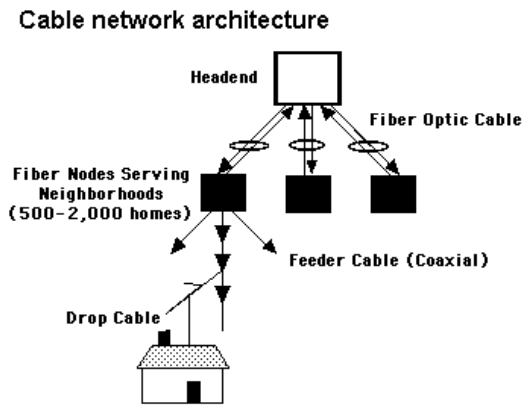

*Figure 1. A typical cable network architecture [1]*

Briefly, the requirements for a domestic network are the following:
- simple and affordable;
- fully compatible with the existing analog TV equipment;
- having the possibility to be expanded to Digital technology without re-cabling or other major modifications;
- extended capabilities to add data services (Internet access, IP- telephony)
- open to future digital TV technology (DVB-C, OpenCable and oth.)

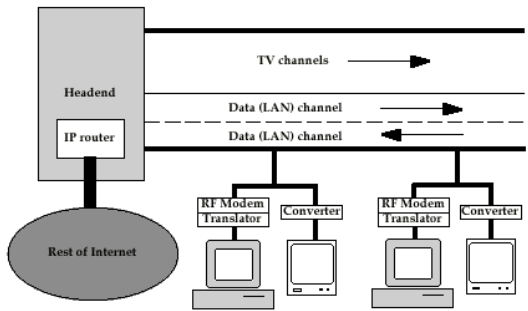

*Figure 2. Standard hybrid solution for data/multimedia distribution [3]*

Cable providers are offering today both TV and data services (including VoIP) using a technology presented in figure 2. This architecture has many advantages (a reliable technology, relative low prices, a comprehensive set of services). But there are some disadvantages:
- the necessity to use only "cable approved" equipment in home network architecture
- the difficulty to add non-standard devices in the system

Reference [1] describes a typical view of cable network architecture and administration. This technology is known as CableHome.

CableHome specifications are intended to provide also being very new, it will take few years until compatible equipment could be affordable for homes. This is the main reason that leads us to investigate different options. One of such options, investigated in our research, is described in the following paragraph.

## II. MULTILET NETWORK ARCHITECTURE
*General Architectural Concepts*

The Multilet system is designed for broadband access over TV cables and utilizes the fact that the TV signal and the 10BASE-T Ethernet signals use different parts of the frequency spectrum. The figure 4 shows how a pair of diplex filters (which separate the data and TV signals) can be used for sending both data and TV on the same cable.

Multilet could be viewed as a cable saving device, as no new cabling needed since an existing CATV cable can also be used for Ethernet. Multilet was designed to solve "the last stairway problem", i.e. how to distribute the data signals within a building. The system is quite neutral how to get to

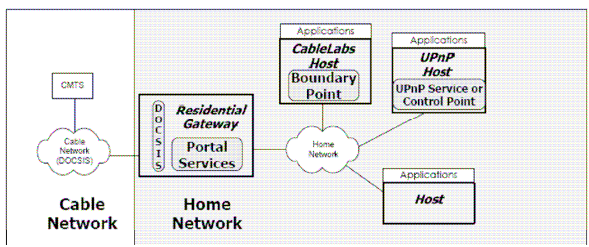

*Figure 3. Logical Reference Architecture of the CableHome system [1]*

Internet Protocol (IP) - based architecture for managed home- networked services on the cable network through a DOCSIS cable modem. The CableHome architecture (figure 3) accommodates any physical and link layer home network technology that supports the transport of IP packets. This layer 1 and 2 independent architecture enables cable operators to provide services to a wide range of home networking environments. The CableHome architecture provides a defined set of requirements that support practically the whole range of services that can be delivered over cable. In order to ensure wide adoption and ease of use of this specification, CableHome technical specifications are aligned with well-known industry standards, as well as other CableLabs projects. CableHome allows the use of the existing cable operators' system infrastructure, but also provides a acceptable transition path for the deployment of CableHome over older systems. The CableHome architecture provides support for existing and future IP-based services into the home.

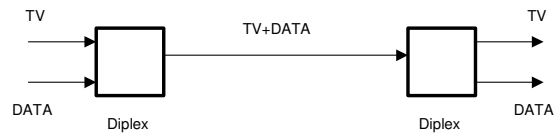

*Figure 4. Main principle of Multilet technology*

This architecture seems to have a promising future, but the building and any available technique may be used:

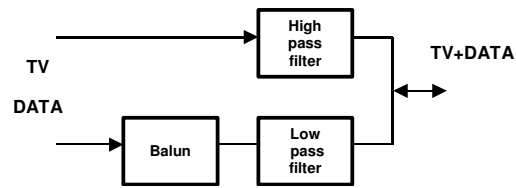

*Figure 5. Block schematic of Multilet combining filter*

leased lines, wireless etc.

Once the data signal has entered the building, it is combined with the CATV-signals, using a passive "combining filter" (CFU, Combining Filter Unit). The principle of combining TV and data is briefly described in figure 5.

Inside each apartment, the signals are separated again using a special wall outlet (WOG, Wall Outlet Generic). In Multilet Cascade, a system where several apartments may share the same cable, a wall adapter (WAC, Wall Adapter Cascade) is needed for those want to get Internet access.

The frequency spectrum created using this bank of filters separates the TV-RF spectrum from Ethernet 10Base-T baseband spectrum. This allocation is described in figure 6.

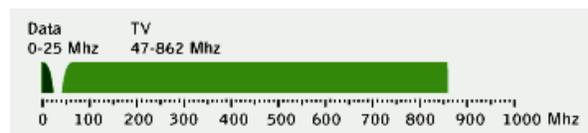

*Figure 6. Frequency allocation in Multilet technology*

This allocation lies on a simple analysis of 10BaseT Ethernet. For 10Base-T Ethernet the fundamental frequency will be between 5 MHz (alternating ones/zeros) and 10 MHz (all ones/all zeros). The energy spectrum of a packetized Ethernet signal using Manchester encoding at 10 Mb/s is concentrated under 30 MHz, with signal energy down to (but not including) DC. For 10baseT, though, the energy is low at lower frequencies, because signal is separated through transformers. The maximum energy is around frequencies between 5-20 MHz (frequencies 5 MHz and 10 MHz being the strongest components). For example in a 10 Mbps Ethernet LAN, the preamble sequence encodes to a 5 MHz square wave.

The following Multilet generation technology, allowing 100Mbit/s access using the same principle, is already available ([4]) but the compatibility with existing TV technology is no longer guaranteed.

*Basic configurations*

There are several different kinds of cable-TV networks (described in figure 7). The most important distinction for Multilet is the cabling topology, i.e. how the cables are connected. The Multilet system exists in two basic variants, Multilet Star, for star-shaped cable TV systems, where each home has its individual cable, and Multilet Cascade, where a number of homes share a single cable.

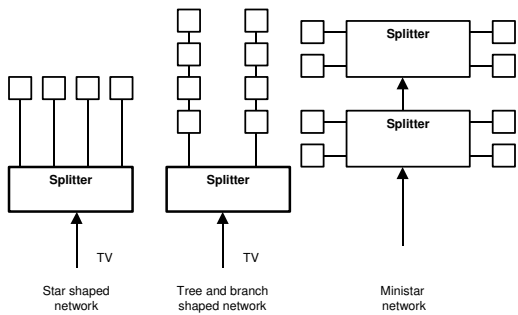

*Figure 7. Basic configurations in TV networks and Multilet technology*

### III. EVALUATION PROCEDURE

*System configuration*

We have built a star shaped system to evaluate the possibilities of Multilet technology. The system configuration is described in figure 8.

Multilet requires that the Ethernet switches connected match the Ethernet standards to operate fairly. There seems to exist some switches that do not follow the Ethernet specification fully. If a previously untested switch is to be used, it is advisable to test it first in the lab, to verify that it works properly ([2]).

There is one more very important thing to consider when using Multilet between buildings: The CFUs support only 10 Mb/s Ethernet. If two 100 Mb/s switches are used in the configuration described above, they will first negotiate the speed using 10 Mb/s, then agreeing on using 100 Mb/s, and the communication probably will stop. The simplest solution to this, is to make at least one of the switches insist on 10 Mb/s Ethernet. It is possible to achieve this in three different ways:
• Install a pure 10 Mb/s Ethernet switch on either side.
• Install a supervised (i.e. remotely controlled) 10/100 Mb/s Ethernet switch, and set the port to 10 Mb/s
• Use a 10 Mb/s hub, and transfer the traffic through it.

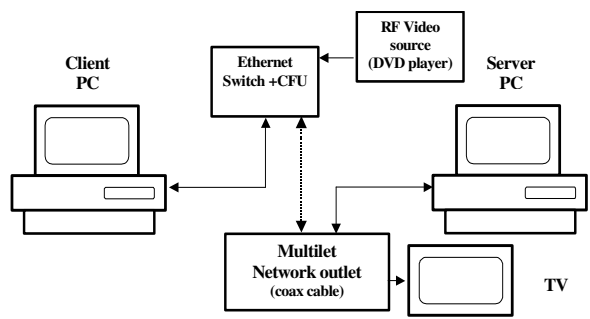

*Figure 8. Test system architecture*

As a practical solution it is sufficient to restrict the speed to 10 Mb/s at just one end of the connection, as the other side will adjust this speed automatically.

*Software tools*

Our goal was to test the behavior of the "new" Ethernet (over Multilet) network. Our interest was to verify the network capabilities in two extreme situations: regular file transfer and multimedia information streaming.

For connections and file transfer we used normal tools (utilities) present in each PC (ping, ifconfig, etc.). In addition, NetLimiter offered information about the quantitative parameters of the connection.

For multimedia streaming, our previous experience in this field, described in [5], was extremely useful.

The main tool used both in server and client was the well-known VLC Media Player ([6]).

### IV. RESULTS

*File transfer*

In a commonly used environment (Win XP), the connection is established automatically, like in a "normal" Ethernet connection.

The screen obtained after connection setup in a client PC is presented in figure 9. The client and the server seem to ignore the Multilet presence and are treating the connection as a "normal" Ethernet connection.

To test the parameters of the connection, like the maximal speed, we used NetMonitor. The monitor screen is presented in figure 10.

The maximum download speed seems to be at the limit of a normal 10Mb Ethernet. The value is around 850MB/s. An important fact that must be taken into account is the lack of the full duplex mode, due to the fundamental proprieties of Multilet. This makes impossible the simultaneous high-speed transfer of information in both directions.

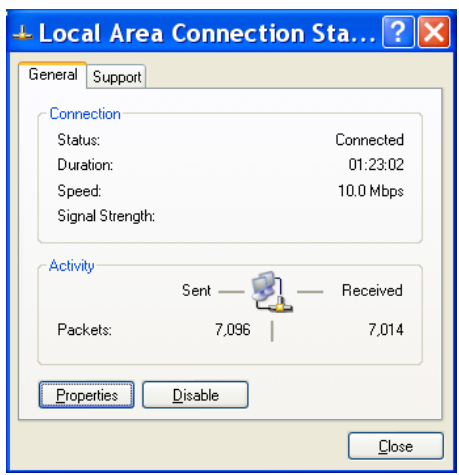

*Figure 9. Connection indication*

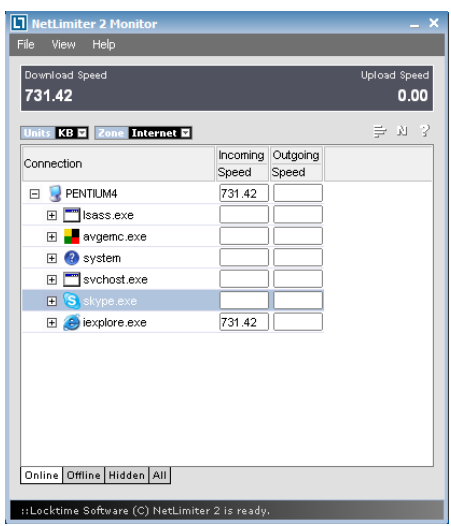

*Figure 10. NetLimiter monitor screen*

*Multimedia streaming*

This type of file transfer is extremely significant for our research, involving entertainment distribution in a normal home. Our question is if this "Multilet cabled house" could be suitable for complete digitalization of entertainment. Both the server and client parts used a well-known program called VLC media Player, developed in VideoLAN project. The main screen of the program is presented in figure 10. The VideoLAN project targets multimedia streaming of MPEG-1, MPEG-2, MPEG-4 and DivX files, DVDs, digital satellite channels, digital terrestrial television channels and live videos on a high-bandwidth IPv4 or IPv6 network in unicast or multicast under the main OSes. VideoLAN also features a multiple platform multimedia player, VLC, which can be used to read the stream from the network or display video read locally on the computer under all GNU/Linux flavours, all BSD versions, Windows, Mac OS X, BeOS, Solaris, QNX, Familiar Linux.

The later feature is one of the most important points for VideoLan, being in fact the only free software able to handle different video and audio formats on different platforms and supporting even "open" architectures like ARM and MIPS based devices. This could be an advantage if integrated digital home entertainment will be implemented in a more efficient way, using embedded technology. VideoLan supports the functionality offered by the IVTV driver, taking advantage of hardware encoding on a growing number of cards. It is basically made out of two software components: VideoLanClient (VLC) and VideoLanServer (VLS). The way VideoLan distributes functionality among its nodes is therefore flexible and permits to easily build streaming configurations to distribute real-time audio/video data.

We used the same software in our previous research [5].

One computer was set as a "video server" and the second as a client. Both used VLC media player for streaming.

In figure 12 it is possible to see the playback process of a MPEG file via Multilet network. The resolution of the transferred image was set at SDTV values (720x576 pixels). Accepting lower resolutions opens the possibility to handle multiple streams.

The speed is excellent for one stream (one channel) and satisfactory for two streams. If we launch simultaneously more than three playbacks the limitations are visible. It is not sure that the limitations are a characteristic of the operating system's networking performance, but we didn't find specific works exploring this direction. In important issue is the quality of the TV RF modulated services during intensive data streaming. We didn't observe any additional disturbances on the control TV's screen.

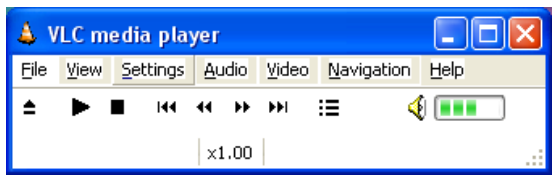

*Figure 11. VLC media player main screen*

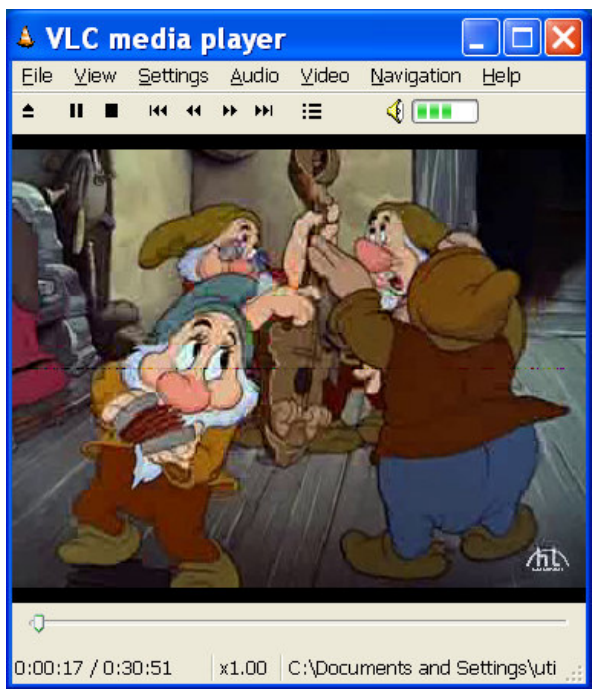

*Figure 12. VLC media player screen during playback*

### V. CONCLUSIONS

This research make possible to evaluate the potential of Multilet in data or multimedia (entertainment) information distribution. The research proved that this kind (multimedia) of application is possible, but with limited features (one or two streams simultaneously).

Much useful is basic Multilet to add data capabilities to conventional (cable TV) connections. The Multilet seems ideal for Internet access, offering about the same performance compared with Cable Modems or ADSL [7].

A second advantage is the fact that the building cabling is minimal avoiding rewiring each apartment, when we need to add a new service.

We didn't test the VoIP capabilities. We didn't test also the capabilities of Multilet using the maximum number of connections (4) offered by our evaluation system. Since the connections are independent, the performance could be the same for all outputs.

We used only the switch supplied by Macab Evaluation Kit. An interesting research would be to test the compatibility for multiple types and brands of switches. Such research reports are present in [4].

The maximum length of the cable connection is also a subject for future investigations. Our connections have a maximum length of 20m. This value is a practical approximation for a normal building.

The speed of the connection is enough for most domestic multimedia and data transmission applications. For the future applications, involving high-speed Internet access, eventually IPTV will be necessary to find additional solutions based on the same concept.

Consequently, a recommendation for the future improvements will be to investigate the possibilities of the new Multilet technology ([4]), providing 100MBps full speed. In this case could be possible to test new IP services (VoIP or IPTV).

Even in this case seems to be possible to have conventional TV (Cable) but with frequencies located in UHF part of the spectrum.

### VI. ACKNOWLEDGEMENTS

This research is included in a CEEX research grant funded by The Romanian National Agency for Research and Technology.

This research was also possible using an evaluation system (star architecture) donated by Macab AB (Sweden).